\documentclass[conference]{IEEEtran}
\IEEEoverridecommandlockouts
\usepackage{cite}
\usepackage{amsmath,amssymb,amsfonts}
\usepackage{algorithmic}
\usepackage{graphicx}
\usepackage{textcomp}
\usepackage{xcolor}
\usepackage{booktabs}
\usepackage{multirow}
\usepackage{tabularx}
\usepackage{url}
\usepackage{tikz}

\def\BibTeX{{\rm B\kern-.05em{\sc i\kern-.025em b}\kern-.08em
    T\kern-.1667em\lower.7ex\hbox{E}\kern-.125emX}}

\newcommand\copyrighttext{%
  \footnotesize This work has been submitted to the IEEE for possible publication. Copyright may be transferred without notice, after which this version may no longer be accessible.}
\newcommand\copyrightnotice{%
\begin{tikzpicture}[remember picture,overlay]
\node[anchor=south,yshift=11pt] at (current page.south) {\fbox{\parbox{\dimexpr\textwidth-\fboxsep-\fboxrule\relax}{\copyrighttext}}};
\end{tikzpicture}%
}

\begin{document}

\title{Energy-Efficient Visual Inspection with FFT-Based CNNs and Adaptive Floating-Point Quantization}

\author{\IEEEauthorblockN{Lukas Krupp\IEEEauthorrefmark{1}, Marco Groß\IEEEauthorrefmark{1}\IEEEauthorrefmark{2}, Michael Graichen\IEEEauthorrefmark{2}, Kim Ulrich\IEEEauthorrefmark{2} and Norbert Wehn\IEEEauthorrefmark{1}}
\IEEEauthorblockA{\IEEEauthorrefmark{1}RPTU University Kaiserslautern-Landau, Kaiserslautern, Germany\\
\IEEEauthorrefmark{2}Wipotec Wiege- und Positioniersysteme GmbH, Kaiserslautern, Germany}
}

\maketitle
\IEEEpubid{\begin{minipage}{\textwidth}
  \copyrightnotice
\end{minipage}} 

\begin{abstract}
This paper investigates reduced-precision floating-point arithmetic for FFT-based CNN inference on an industrial CPU–FPGA platform. We combine FFT-based convolution with adaptive post-training FP8 quantization and evaluate two FPGA-oriented optimization methods: progressive bias adjustment (PBA) within the FFT and layer-wise exponent-bias selection across the CNN. The methods are implemented in a LeNet-5 accelerator using serial radix-$\mathbf{2^2}$ SDF FFT modules and evaluated on an industrial fault detection dataset. Results show that weight scaling outperforms PBA, while layer-wise bias optimization increases the accuracy from 80.33\% to 84.13\% without modifying the datapath width. Compared with CPU-only inference, the FPGA achieves approximately 2.5$\times$ higher energy efficiency.
\end{abstract}


\section{Introduction}
In industrial inspection, defects such as missing items, packaging damage, or product deviations cannot always be reliably detected using traditional image processing methods. Convolutional neural networks (CNNs) have therefore become an established approach for automated inspection, as they can learn task-specific visual features directly from data. In practice, CNN inference is performed mostly at the edge in industrial computers close to the production line. These systems operate continuously under constraints on power consumption, thermal design power (TDP), cooling capacity, and enclosure volume. Minimizing the power and energy required per inspected item is therefore an important design objective, making FPGAs attractive due to their potential for power- and energy-efficient, application-specific acceleration.

Fast Fourier transform (FFT)-based convolution reduces the arithmetic complexity of convolution layers while preserving their mathematical function and is therefore employed in state-of-the-art software libraries such as cuDNN \cite{b1}. 
However, fixed-point arithmetic remains the predominant choice both for FPGA implementations of FFT-based CNNs \cite{b2, b3, b4} and for FFTs in digital signal processing applications \cite{b5}. In contrast, reduced-precision floating-point formats have become an important driver of performance and energy efficiency in deep learning accelerators \cite{b6}. Their application to FFT operations, and particularly to FFT-based CNN acceleration, has nevertheless received little attention.

This work therefore investigates how custom low-bitwidth floating-point formats affect FFT-based CNN inference and explores FPGA-oriented techniques for improving numerical robustness.
We present the following key contributions:
\begin{itemize}
    \item An experimental investigation of low-bitwidth floating-point arithmetic for FFT-based CNN inference on FPGA.
    \item FFT-level and layer-wise data format optimization methods for improving post-training quantization accuracy.
    \item An evaluation on an industrial X-ray inspection task and dataset using a CPU--FPGA platform deployed in real production environments.
\end{itemize}

To the best of our knowledge, this is the first experimental study of custom reduced-precision floating-point formats for FFT-based CNN acceleration on FPGAs.

\section{Methodology}
\begin{figure}[htbp]
\centerline{\includegraphics[width=0.4\textwidth]{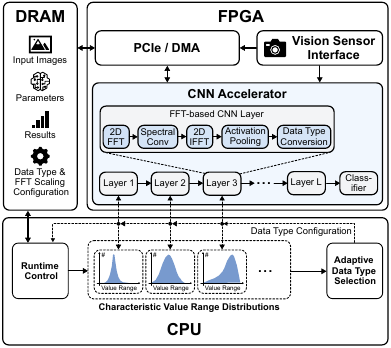}}
\caption{Overview of the CPU--FPGA system architecture.}
\label{fig:system_architecture}
\end{figure}

Figure~\ref{fig:system_architecture} shows the proposed CPU--FPGA architecture. Industrial inspection systems often already combine a commercial off-the-shelf (COTS) CPU with an FPGA. The CPU controls image acquisition and system operation, while the FPGA implements sensor, PCIe, and machine interfaces. In this work, we additionally use the FPGA to offload CNN inference to a dedicated accelerator, with input data, parameters, and results exchanged through DRAM.

The CNN is implemented using FFT-based convolutions and reduced-precision floating-point arithmetic. We optimize the numerical format at two levels. At FFT level, we compare \textit{Progressive Bias Adjustment} (PBA) with static weight scaling. PBA compensates for the stage-wise growth of FFT values by decrementing the exponent bias by one per butterfly stage. Weight scaling instead applies the inverse FFT factor $\frac{1}{N}$ to the pre-transformed weights before quantization.

At CNN level, the mantissa and exponent widths remain fixed, while the exponent bias is selected individually for each layer based on its observed value range. Since the bias can be stored as a configuration value, it can be changed at runtime without modifying the datapath width. Format converters between adjacent layers apply the bias changes. The layer-wise bias configuration is determined by Bayesian optimization, which evaluates candidate bias combinations on a small calibration subset collected during operation.

\section{Implementation \& Experimental Setup}
The prototype is implemented on an industrial computer used in real production lines, comprising an Intel Core i7 CPU and a Xilinx XC7A200T FPGA. The accelerator implements a LeNet-5 CNN with FFT-based convolutions.

The 2D-FFT and IFFT operations are realized in SystemVerilog using serial radix-$2^2$ single-path delay-feedback (SDF) modules. Their serial architecture requires only one butterfly unit per FFT stage, resulting in low area utilization. This is important because the FPGA is mainly used for its original functions and not dedicated exclusively to CNN acceleration.

\begin{figure}[htbp]
\centerline{\includegraphics[width=0.25\textwidth]{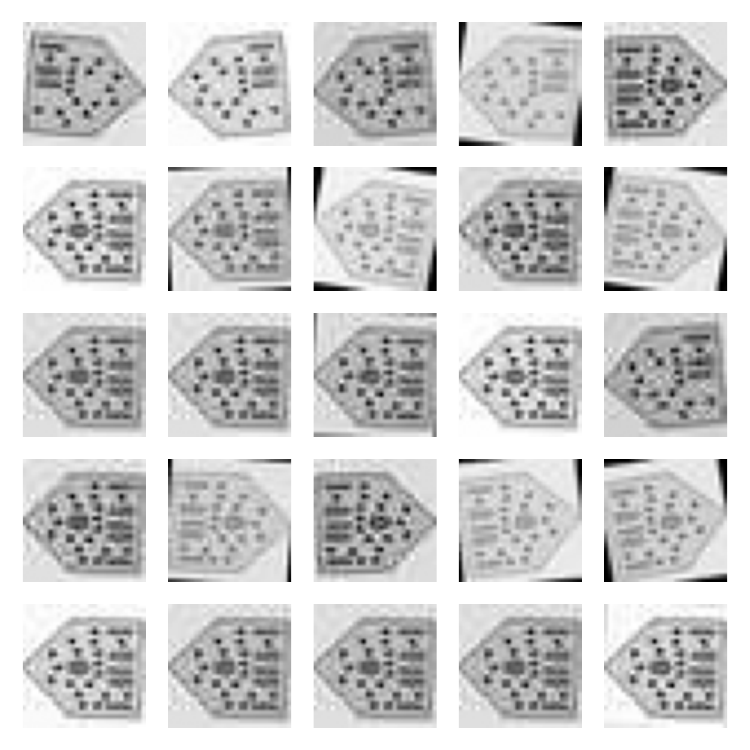}}
\caption{Samples from the fault detection dataset.}
\label{fig:dataset_examples}
\end{figure}

Experiments are conducted on an industrial fault-detection dataset containing 3,208 X-ray images of Advent calendars. The images shown in Fig.~\ref{fig:dataset_examples} are labeled as faulty or fault-free.

Post-training quantization is used to map already trained networks to different reduced-precision formats without retraining. This reduces deployment effort and supports runtime-adaptive format configurations for changing inspection tasks. The mantissa and exponent bitwidths are globally fixed to three and four bits, respectively. This floating-point format denoted as FP8 (E4M3) has been shown to provide suitable accuracy for CNN forward passes in prior work \cite{b7}.

\section{Results}
Table~\ref{tab:accuracy} summarizes the classification results. Weight scaling consistently outperforms PBA. Layer-wise bias optimization improves accuracy by 1.75\% for PBA and by 3.80\% for weight scaling, reaching a maximum reduced-precision accuracy of 84.13\%. This indicates that layer-wise bias search is a low-cost additional tuning step for reduced-precision floating-point formats, since the resulting biases can be configured at runtime without modifying the FPGA datapath. Nevertheless, a substantial gap to FP32 remains. This remaining accuracy gap may be reduced by increasing the mantissa width.

\begin{table}[h]
    \centering
    \caption{Classification accuracy for different quantization methods.}
    \label{tab:accuracy}
    \setlength{\tabcolsep}{0pt}
    \begin{tabular*}{\columnwidth}{@{\extracolsep{\fill}}lcc@{}}
        \toprule
        \textbf{FFT Scaling / Arithmetic} &
        \textbf{Layer-wise Bias} &
        \textbf{Accuracy} \\
        \midrule
        Standard / FP32 (Reference)
            & --  & 97.94\% \\
        \midrule
        \multirow{2}{*}{PBA / FP8 (E4M3)}
            & No  & 72.91\% \\
            & Yes & 74.66\% \\
        \midrule
        \multirow{2}{*}{Weight Scaling / FP8 (E4M3)}
            & No  & 80.33\% \\
            & Yes & \textbf{84.13\%} \\
        \bottomrule
    \end{tabular*}
\end{table}

The accelerator fits alongside the existing logic on the FPGA and therefore fulfills the area requirements. Compared with CPU inference, the FPGA has a higher latency of 1.91~ms versus 32.90~$\mu$s, but reduces power consumption from 53.73~W to 0.37~W. This corresponds to approximately 0.70~mJ per FPGA inference and 1.77~mJ per CPU inference, yielding an energy-efficiency improvement of $\sim2.5\times$, including CPU power, PCIe, and DMA transfers for the FPGA case.

\section{Conclusion}
This work investigated low-precision floating-point arithmetic for FFT-based CNN inference on FPGAs. The proposed PBA method did not outperform static weight scaling. This is likely caused by its worst-case assumption of a factor-two growth per FFT stage, which reduces the exponent bias by one regardless of the actual value distribution. A distribution-aware bias selection within the FFT may therefore provide better results. In contrast, layer-wise bias optimization substantially improved classification accuracy without hardware overhead. Overall, the results indicate that combining FFT-based convolution with adaptive floating-point quantization is a promising direction for energy-efficient CNN acceleration.

Future work will evaluate larger mantissa widths to increase accuracy and extend the optimization space to exponent and mantissa bitwidths. Partial reconfiguration of modern FPGAs enables adaptation of the datapath bitwidth at runtime.

\end{document}